\documentclass[twocolumn,amsmath,amssymb,aps,superscriptaddress]{revtex4-1}

\usepackage{graphicx}
\usepackage{dcolumn}
\usepackage{bm}
\usepackage[mathlines]{lineno}
\usepackage{braket}

\begin{document}

\preprint{}

\title{Measurement of the multi-TeV neutrino cross section with IceCube using Earth absorption}

\affiliation{III. Physikalisches Institut, RWTH Aachen University, D-52056 Aachen, Germany}
\affiliation{Department of Physics, University of Adelaide, Adelaide, 5005, Australia}
\affiliation{Dept.~of Physics and Astronomy, University of Alaska Anchorage, 3211 Providence Dr., Anchorage, AK 99508, USA}
\affiliation{Dept.~of Physics, University of Texas at Arlington, 502 Yates St., Science Hall Rm 108, Box 19059, Arlington, TX 76019, USA}
\affiliation{CTSPS, Clark-Atlanta University, Atlanta, GA 30314, USA}
\affiliation{School of Physics and Center for Relativistic Astrophysics, Georgia Institute of Technology, Atlanta, GA 30332, USA}
\affiliation{Dept.~of Physics, Southern University, Baton Rouge, LA 70813, USA}
\affiliation{Dept.~of Physics, University of California, Berkeley, CA 94720, USA}
\affiliation{Lawrence Berkeley National Laboratory, Berkeley, CA 94720, USA}
\affiliation{Institut f\"ur Physik, Humboldt-Universit\"at zu Berlin, D-12489 Berlin, Germany}
\affiliation{Fakult\"at f\"ur Physik \& Astronomie, Ruhr-Universit\"at Bochum, D-44780 Bochum, Germany}
\affiliation{Universit\'e Libre de Bruxelles, Science Faculty CP230, B-1050 Brussels, Belgium}
\affiliation{Vrije Universiteit Brussel (VUB), Dienst ELEM, B-1050 Brussels, Belgium}
\affiliation{Dept.~of Physics, Massachusetts Institute of Technology, Cambridge, MA 02139, USA}
\affiliation{Dept. of Physics and Institute for Global Prominent Research, Chiba University, Chiba 263-8522, Japan}
\affiliation{Dept.~of Physics and Astronomy, University of Canterbury, Private Bag 4800, Christchurch, New Zealand}
\affiliation{Dept.~of Physics, University of Maryland, College Park, MD 20742, USA}
\affiliation{Dept.~of Physics and Center for Cosmology and Astro-Particle Physics, Ohio State University, Columbus, OH 43210, USA}
\affiliation{Dept.~of Astronomy, Ohio State University, Columbus, OH 43210, USA}
\affiliation{Niels Bohr Institute, University of Copenhagen, DK-2100 Copenhagen, Denmark}
\affiliation{Dept.~of Physics, TU Dortmund University, D-44221 Dortmund, Germany}
\affiliation{Dept.~of Physics and Astronomy, Michigan State University, East Lansing, MI 48824, USA}
\affiliation{Dept.~of Physics, University of Alberta, Edmonton, Alberta, Canada T6G 2E1}
\affiliation{Erlangen Centre for Astroparticle Physics, Friedrich-Alexander-Universit\"at Erlangen-N\"urnberg, D-91058 Erlangen, Germany}
\affiliation{D\'epartement de physique nucl\'eaire et corpusculaire, Universit\'e de Gen\`eve, CH-1211 Gen\`eve, Switzerland}
\affiliation{Dept.~of Physics and Astronomy, University of Gent, B-9000 Gent, Belgium}
\affiliation{Dept.~of Physics and Astronomy, University of California, Irvine, CA 92697, USA}
\affiliation{Dept.~of Physics and Astronomy, University of Kansas, Lawrence, KS 66045, USA}
\affiliation{SNOLAB, 1039 Regional Road 24, Creighton Mine 9, Lively, ON, Canada P3Y 1N2}
\affiliation{Dept.~of Astronomy, University of Wisconsin, Madison, WI 53706, USA}
\affiliation{Dept.~of Physics and Wisconsin IceCube Particle Astrophysics Center, University of Wisconsin, Madison, WI 53706, USA}
\affiliation{Institute of Physics, University of Mainz, Staudinger Weg 7, D-55099 Mainz, Germany}
\affiliation{Department of Physics, Marquette University, Milwaukee, WI, 53201, USA}
\affiliation{Universit\'e de Mons, 7000 Mons, Belgium}
\affiliation{Physik-department, Technische Universit\"at M\"unchen, D-85748 Garching, Germany}
\affiliation{Institut f\"ur Kernphysik, Westf\"alische Wilhelms-Universit\"at M\"unster, D-48149 M\"unster, Germany}
\affiliation{Bartol Research Institute and Dept.~of Physics and Astronomy, University of Delaware, Newark, DE 19716, USA}
\affiliation{Dept.~of Physics, Yale University, New Haven, CT 06520, USA}
\affiliation{Dept.~of Physics, University of Oxford, 1 Keble Road, Oxford OX1 3NP, UK}
\affiliation{Dept.~of Physics, Drexel University, 3141 Chestnut Street, Philadelphia, PA 19104, USA}
\affiliation{Physics Department, South Dakota School of Mines and Technology, Rapid City, SD 57701, USA}
\affiliation{Dept.~of Physics, University of Wisconsin, River Falls, WI 54022, USA}
\affiliation{Dept.~of Physics and Astronomy, University of Rochester, Rochester, NY 14627, USA}
\affiliation{Oskar Klein Centre and Dept.~of Physics, Stockholm University, SE-10691 Stockholm, Sweden}
\affiliation{Dept.~of Physics and Astronomy, Stony Brook University, Stony Brook, NY 11794-3800, USA}
\affiliation{Dept.~of Physics, Sungkyunkwan University, Suwon 440-746, Korea}
\affiliation{Dept.~of Physics and Astronomy, University of Alabama, Tuscaloosa, AL 35487, USA}
\affiliation{Dept.~of Astronomy and Astrophysics, Pennsylvania State University, University Park, PA 16802, USA}
\affiliation{Dept.~of Physics, Pennsylvania State University, University Park, PA 16802, USA}
\affiliation{Dept.~of Physics and Astronomy, Uppsala University, Box 516, S-75120 Uppsala, Sweden}
\affiliation{Dept.~of Physics, University of Wuppertal, D-42119 Wuppertal, Germany}
\affiliation{DESY, D-15738 Zeuthen, Germany}

\author{M.~G.~Aartsen}
\affiliation{Department of Physics, University of Adelaide, Adelaide, 5005, Australia}
\author{M.~Ackermann}
\affiliation{DESY, D-15738 Zeuthen, Germany}
\author{J.~Adams}
\affiliation{Dept.~of Physics and Astronomy, University of Canterbury, Private Bag 4800, Christchurch, New Zealand}
\author{J.~A.~Aguilar}
\affiliation{Universit\'e Libre de Bruxelles, Science Faculty CP230, B-1050 Brussels, Belgium}
\author{M.~Ahlers}
\affiliation{Niels Bohr Institute, University of Copenhagen, DK-2100 Copenhagen, Denmark}
\author{M.~Ahrens}
\affiliation{Oskar Klein Centre and Dept.~of Physics, Stockholm University, SE-10691 Stockholm, Sweden}
\author{I.~Al~Samarai}
\affiliation{D\'epartement de physique nucl\'eaire et corpusculaire, Universit\'e de Gen\`eve, CH-1211 Gen\`eve, Switzerland}
\author{D.~Altmann}
\affiliation{Erlangen Centre for Astroparticle Physics, Friedrich-Alexander-Universit\"at Erlangen-N\"urnberg, D-91058 Erlangen, Germany}
\author{K.~Andeen}
\affiliation{Department of Physics, Marquette University, Milwaukee, WI, 53201, USA}
\author{T.~Anderson}
\affiliation{Dept.~of Physics, Pennsylvania State University, University Park, PA 16802, USA}
% ...

\date{today}
\collaboration{The IceCube Collaboration}
\noaffiliation

\maketitle

{\bf Neutrinos interact only very weakly, so they are extremely penetrating.   However, the theoretical neutrino-nucleon interaction cross section rises with energy such that, at energies above 40 TeV, neutrinos are expected to be absorbed as they pass through the Earth.   Experimentally, the cross section has been measured only at the relatively low energies (below 400 GeV) available at neutrino beams from accelerators \cite{Agashe:2014kda, Formaggio:2013kya}.  Here we report  the first measurement of neutrino absorption in the Earth, using a sample of 10,784 energetic upward-going neutrino-induced muons observed with the IceCube Neutrino Observatory.  The flux of high-energy neutrinos transiting long paths through the Earth is attenuated compared to a reference sample that follows shorter trajectories through the Earth.  Using a fit to the two-dimensional distribution of  muon energy and zenith angle, we determine the cross section for neutrino energies between 6.3 TeV and 980 TeV, more than an order of magnitude higher in energy than previous measurements.  The measured cross section is $1.30^{+0.21}_{-0.19}$ (stat.) $^{+0.39}_{-0.43}$ (syst.) times the prediction of the Standard Model \cite{CooperSarkar:2011pa}, consistent with the expectation for charged and neutral current interactions.   We do not observe a dramatic increase in the cross section, expected in some speculative models, including those invoking new compact dimensions \cite{AlvarezMuniz:2002ga} or the production of leptoquarks \cite{Romero:2009vu}. 
}

\begin{figure*}
 \includegraphics[width=0.8\textwidth]{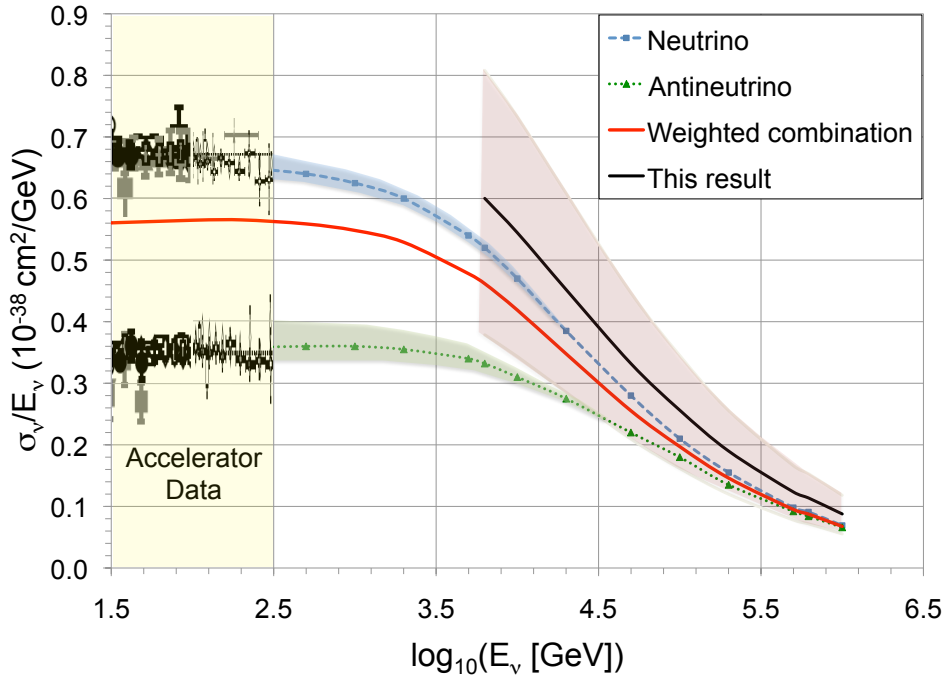} 
\caption{{\bf Neutrino cross section measurements.} Compilation of neutrino charged current cross section measurements, divided by neutrino energy, from accelerator experiments, from \cite{Agashe:2014kda} and the current result.   
The blue and green lines are the Standard Model predictions for $\nu_\mu$ and $\overline\nu_\mu$ respectively, with the uncertainties on the deep inelastic cross sections shown by the shaded bands \cite{CooperSarkar:2011pa}.  The red line is for the expected mixture of $\nu_\mu$ and $\overline\nu_\mu$ in the IceCube sample.  The black line shows the current result, assuming that the charged and neutral current cross sections vary in proportion, and that the ratio between the actual cross section and the Standard Model prediction does not depend on energy.  The pink band shows the total $1\sigma$ (statistical plus systematic) uncertainty.   The cross section rises linearly with energy up to about 3 TeV, but then the increase moderates, to roughly as $E_\nu^{0.3}$, due to the finite $W^\pm$ and $Z^0$ masses.
\label{fig:summary}
}
\label{fig:result}
\end{figure*}

The cross section for neutrino interactions with matter is very small. Neutrinos are popularly regarded as particles that will go through anything \cite{anything}.   However, the neutrino-nucleon cross section is expected to increase with energy.  
To date, as Fig. \ref{fig:summary} shows, the cross section has only been measured up to a neutrino energy of 370 GeV, limited by the available accelerator neutrino beams  \cite{Agashe:2014kda}.   In this range, the cross section rises linearly with energy.   

In the Standard Model of particle physics, neutrinos interact with quarks through charged current and neutral current interactions, mediated by $W^\pm$ and $Z^0$ bosons, respectively.     At neutrino energies above 10 TeV, the finite $W^\pm$ and $Z^0$ masses are expected to moderate the increase in cross section, leading to a slower rise at higher energies.  The cross sections also reflect the densities of partons
(quarks and gluons) within the nuclear targets.  Accelerator neutrino experiments have mainly probed the densities of partons with Bjorken$-x$ values (the fraction of the total nucleon momentum carried by a quark or gluon)  above about 0.1.  In this $x$ range, there are more quarks than antiquarks, and so the antineutrino cross section is about half that of the neutrino.  Higher-energy experiments probe lower Bjorken$-x$ values, where sea quarks predominate, and the difference between the neutrino and antineutrino cross sections is reduced.    
 
 At high energies, new beyond-Standard-Model processes may appear.  Some theories invoke new spatial dimensions which are curled-up on a distance scale $r$.  At momentum transfers comparable to $\hbar c/r$, the neutrino cross section rises dramatically \cite{AlvarezMuniz:2002ga, Connolly:2011vc}.  In some grand unified or technicolor theories, leptoquarks may couple to both quarks and leptons  {\it e. g. } a second-generation leptoquark couples to both muon neutrinos and quarks.  The cross section rises dramatically at neutrino-quark center-of-mass energies corresponding to the mass of the leptoquark \cite{Romero:2009vu}.

\begin{figure*}
\includegraphics[width=0.95\textwidth]{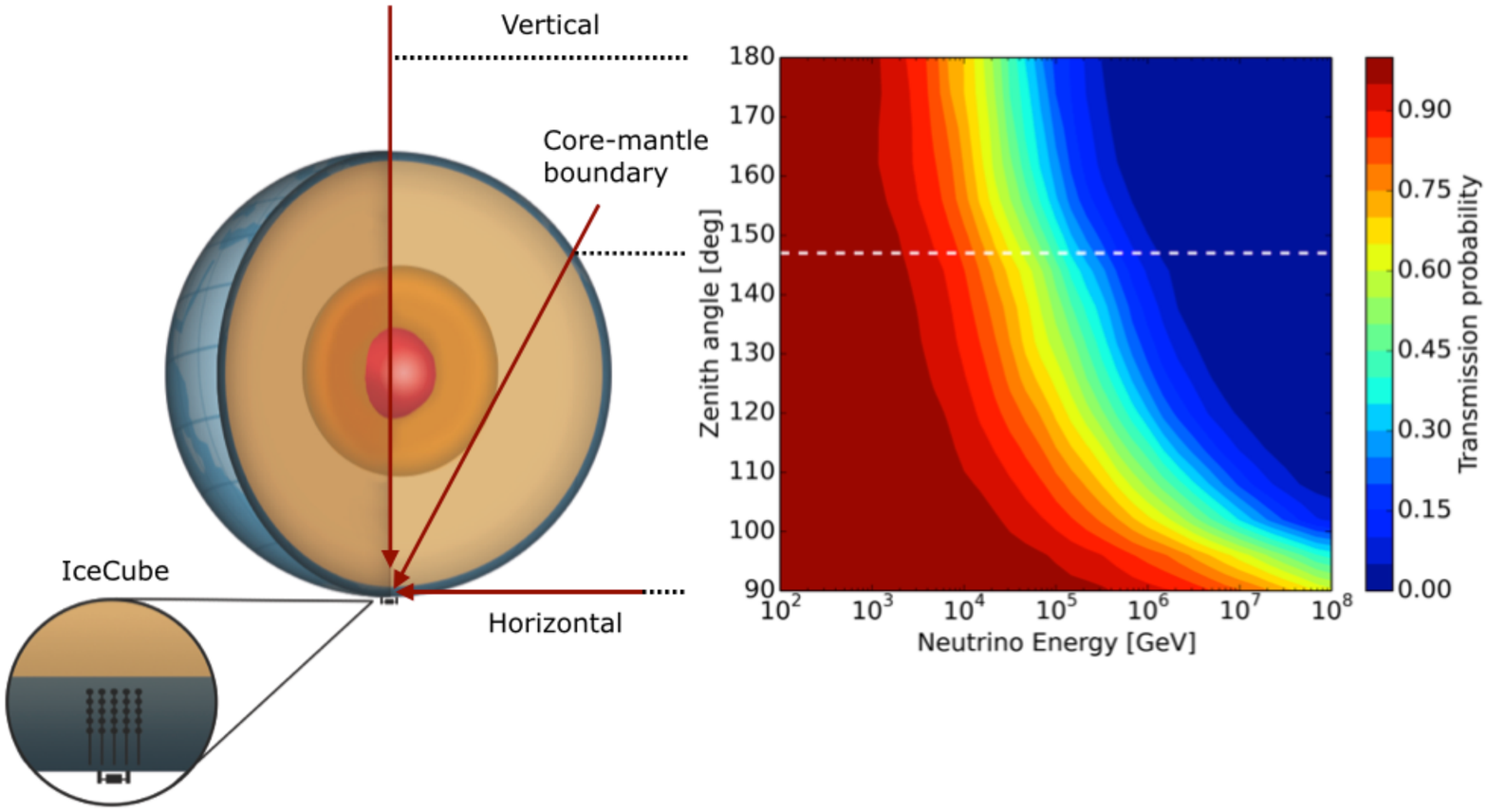}
\caption{{\bf Neutrino absorption in the Earth.} Neutrino absorption is observed by measuring how the neutrino energy spectrum changes with the zenith angle.  High-energy neutrinos transiting deep through the Earth are absorbed, while low-energy neutrinos are not.  Neutrinos from just below the horizon provide a nearly absorption-free baseline at all relevant energies.  The right-hand side shows the transmission probability predicted by the Standard Model for neutrinos to transit the Earth as a function of energy and zenith angle.  Neutral current interactions, which occur about 1/3 of the time are included.  When a neutral current interaction occurs, a neutrino is replaced with a neutrino with lower energy.  The horizontal  dotted white line shows a neutrino trajectory (and zenith angle) that just passes through the core-mantle boundary.}
\label{fig:principle}
\end{figure*}

This measurement uses naturally occurring atmospheric and astrophysical neutrinos to extend cross section measurements to multi-TeV energies by observing, for the first time, neutrino absorption in the Earth.    Figure \ref{fig:principle} shows the principle of the measurement.  Atmospheric neutrinos, produced by cosmic-ray air showers below the Earth's horizon, are the dominant source of neutrinos for this analysis.  Astrophysical neutrinos produced by distant sources make a small contribution at the highest energies \cite{Aartsen:2015knd}.  High-energy neutrinos that deeply traverse the Earth are absorbed, while near-horizontal neutrinos provide an essentially absorption-free reference.   Details are available in Ref. \cite{Sandydissertation}.   The contribution of atmospheric neutrino oscillations is negligible at TeV energies and is not included here.

The idea of studying neutrino absorption in the Earth dates back to 1974 \cite{Volkova}, although most of the early papers on the subject propose using absorption to probe the Earth's interior \cite{Wilson:1983an}.   However,  the density uncertainty for long paths through the Earth is only 1-2\% \cite{PREM,accuracy}; this leads to a less than 1\% systematic uncertainty on the cross section measurement, below the total uncertainty on the cross section.    The early papers on the subject envisioned using accelerator neutrinos for Earth tomography; the idea of using natural (astrophysical and/or atmospheric) neutrinos came later \cite{Hank,GonzalezGarcia:2007gg}.

Neutrino absorption increases with neutrino energy, such that for 40 TeV neutrinos, the Earth's diameter is one absorption length.   By  observing the change in the angular distribution of Earth-transiting neutrinos with increasing neutrino energy, one can measure the increasing absorption and, from that, determine the cross section.  

The analysis uses data collected during 2009 and 2010 when the IceCube detector \cite{Aartsen:2016nxy} consisted of 79 vertical strings  \cite{Halzen:2010yj}, each supporting 60 optical sensors (Digital Optical Modules, DOMs \cite{Abbasi:2008aa}). The strings are arranged in a 125 m triangular grid, with the sensors deployed  at 17 m vertical intervals in the Antarctic ice cap at the South Pole, at depths between 1450 m and 2450 m below the surface.   Six of the strings are emplaced in the center of the array, with a smaller string spacing and with their DOMs clustered between 2100 and 2450 m deep.  This is ``DeepCore." 

The DOMs detect Cherenkov light from the charged particles that are produced when neutrinos interact in the ice  surrounding IceCube and the bedrock below.  The 79-string detector recorded about 2,000 events per second. About 99.9999\% of these were downward-going muons produced directly in cosmic-ray air showers above the horizon.     Events were reconstructed with a series of algorithms of increasing accuracy and computational complexity \cite{Aartsen:2015rwa,Weaver}.  At each stage of processing, a set of selections was applied to eliminate background events.  The final sample of 10,784 upward-going (zenith angle greater than $90^\circ$) events has an estimated background of less than 0.1\%.  Almost all of these are mis-reconstructed downward-going muons.
 
The neutrino zenith angles were determined from the reconstructed muon direction.  The typical angular resolution was better than $0.6^\circ$, including the angular difference between the neutrino and muon directions.  This small uncertainty does not impact the result.  The neutrino energies are much less well known because we do not know how far from the detector the interaction occurred, so we do not know how much energy the muon lost before entering the detector.  Therefore, this analysis used the muon energy as determined via the measured specific energy loss ($dE/dx$) of the muons.  To improve the resolution, the muon tracks were divided into 120 m long segments.  The segments with the highest $dE/dx$ values were excluded, and a truncated mean was determined from the remaining segments \cite{Abbasi:2012wht}.  The removal of large stochastic losses leads to better resolution than the untruncated mean.   The muon energy can be determined to roughly a factor of 2.
 
 The cross section  is found by a maximum likelihood fit which compares the data, binned in zenith angle and muon energy, with a model that includes contributions from atmospheric and astrophysical neutrinos.  The cross section enters the fit through an energy and zenith-angle dependent probability for the neutrinos to be absorbed as they pass through the Earth.    The absorption probability depends on the nucleon density integrated along the path through the Earth.  We use the Preliminary Reference Earth Model for the Earth's density \cite{PREM}.  Thanks to seismic wave studies and tight constraints on the Earth's total mass,  the uncertainties in the integrated density are less than a few percent.    

To account for neutral current interactions, where neutrinos lose a  fraction of their energy, the analysis models neutrino transmission through the Earth at each zenith angle in two dimensions: incident neutrino energy and neutrino energy near IceCube.   The fit determined $R$ where $R=\sigma_{\rm meas.}/\sigma_{\rm SM}$, where $\sigma_{\rm SM}$ is the Standard Model cross section from Ref. \cite{CooperSarkar:2011pa}.  That calculation used quark and gluon densities derived from HERA data to find the neutrino and antineutrino cross sections on protons and neutrons, treating the Earth as an isoscalar target.   The estimated uncertainty in the calculation is less than 5\% for the energy range covered by this analysis.   Because it did not include nuclear shadowing, it may overestimate the cross section for heavier elements, such as the iron in the Earth's core.   Experiments with 2-22 GeV neutrinos interacting in iron targets~\cite{Mousseau:2016snl} and 20-300 GeV neutrinos interacting in neon \cite{CooperSarkar:1984eb} did not observe nuclear shadowing, but it may be larger for higher energy neutrinos \cite{Eskola:2016oht}.  

The fitted charged current and neutral current cross sections are assumed to be the same multiple of their Standard Model counterparts, and we ignore nuclear shadowing. The fit procedure is repeated for different cross section values (varying in steps of $\Delta R=0.2$), leading to a parabolic curve of likelihood vs. cross section

The flux model includes conventional atmospheric neutrinos from $\pi^\pm$ and $K^\pm$ decay, prompt atmospheric neutrinos from the decay of charm/bottom hadrons, and astrophysical neutrinos.   The neutrino fluxes and spectra are imperfectly known, and so they were included as nuisance parameters in the fit, with the initial values and Gaussian uncertainties  shown in Tab. I.  Five parameters accounted for the three neutrino fluxes ($\Phi$) and two spectral indices. The other parameters were the $K/\pi$  and $\nu/\overline\nu$ ratios in cosmic-ray air showers, plus one parameter to account for the overall optical efficiency of the IceCube DOMs.

 \begin{figure*}
  \includegraphics[width=0.7\textwidth]{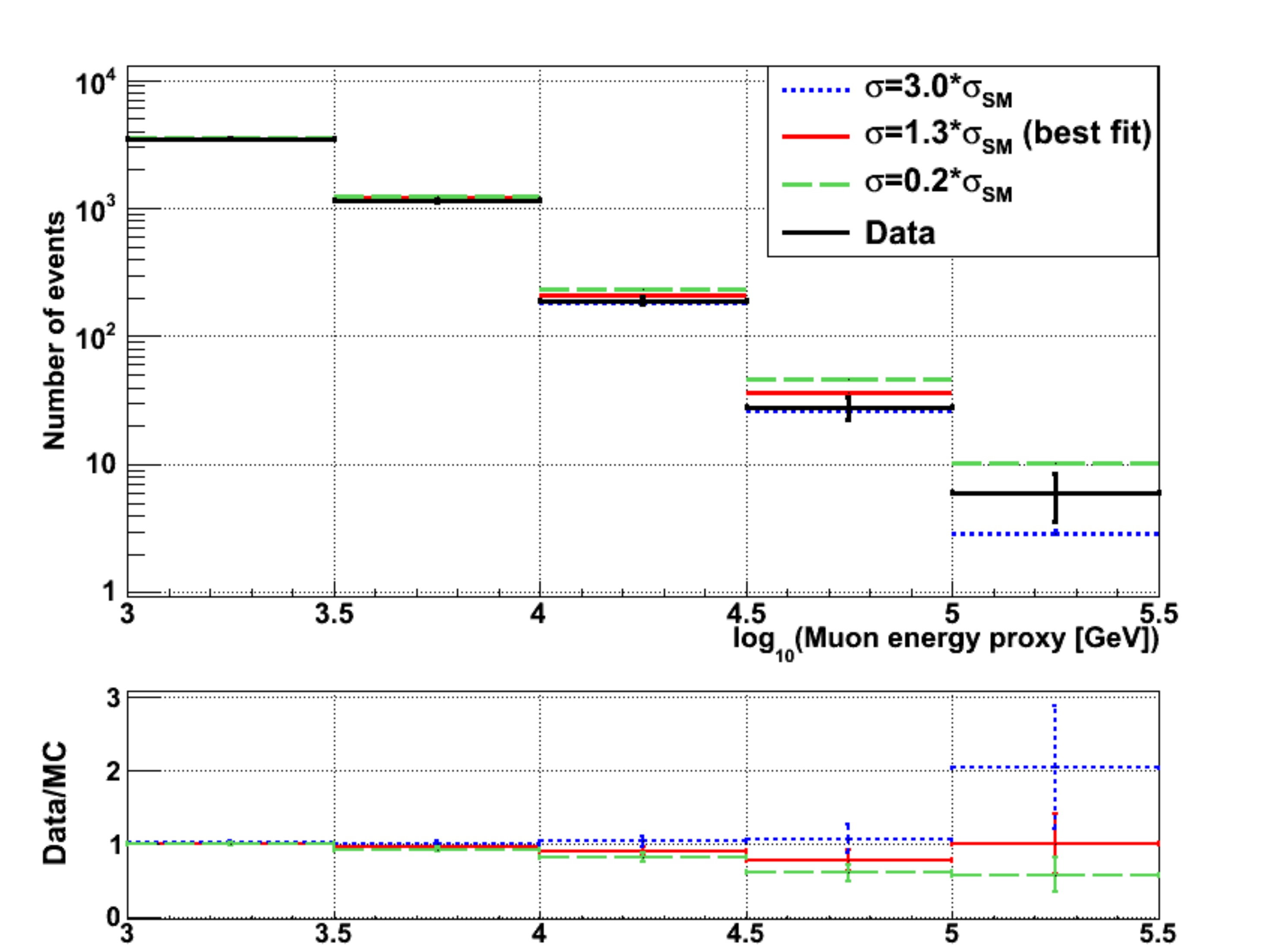}
\caption{{\bf Data compared with expectations for different cross sections.}  Energy spectrum of the data (black points) and the best-fit results (red curve) with the cross sections fixed to 0.2  (green) and 3 times the Standard Model (blue) for events with zenith angles between 110 and 180 degrees, where absorption is significant.  The bottom panel shows the ratios of the data to those three Monte Carlo predictions.  The error bars show the 1$\sigma$ (statistical only) errors.}
\label{fig:spectra}
\end{figure*}

The prior conventional and prompt atmospheric neutrino spectra come from cosmic-ray air-shower simulations, which are tied respectively to lower energy neutrino data \cite{Honda:2006qj} and a color dipole model calculation \cite{Enberg:2008jm}.  The spectra were  modified by us to account for the steepening of the cosmic-ray spectrum at the knee \cite{Aartsen:2016xlq}.   Recent perturbative QCD calculations \cite{Bhattacharya:2015jpa} found a lower prompt flux, but the prompt component has little effect on this analysis, and the fit results are compatible with either calculation and with existing upper limits \cite{Aartsen:2016xlq}.   Finally, the astrophysical spectrum was based on a recent combined fit~\cite{Aartsen:2015knd}.  There is some tension between the spectral index from the combined fit and a newer analysis focused on through-going muon tracks from $\nu_\mu$ \cite{Aartsen:2016xlq}; this tension is treated as a systematic uncertainty due to the uncertain spectral index.

Since past measurements of the neutrino flux were based on the assumption that the Standard Model cross section is correct, this fit uses the product of each flux with the cross section, in order to directly apply constraints from these prior data.  As the cross section rises, the fluxes must drop to preserve the total number of events observed in those experiments.  The fit is thus sensitive to neutrino absorption in the Earth, and not to the total number of observed events.

The fit finds a cross section of $1.30^{+0.30}_{-0.26}$ times the Standard Model.  The uncertainty is a mixture of the statistical uncertainty and the systematic errors from the uncertainties in the nuisance parameters.  We isolate the statistical error by refitting, with the nuisance parameters fixed to their preferred values, and find a statistical error of $^{+0.21}_{-0.19}$. The remainder of the fit error, $^{+0.21}_{-0.18}$ after quadrature subtraction, is attributed to systematic sources in the fit.  

Figure \ref{fig:spectra} compares the muon energy proxy spectrum, for zenith angles between 110 and 180 degrees (where absorption is significant) of the data with three fits: a best-fit result (using the cross section result above) and two comparison fits, with the cross sections fixed to 0.2 and 3.0 times the Standard Model prediction, respectively.   The spectrum steepens noticeably as the cross section increases.  We use the label ``energy proxy"  because of the limited energy resolution.

The other major detector-related uncertainty is due to the optical properties of the ice.  This was studied with separate dedicated simulations,  where the scattering and absorption lengths were varied by $\pm 10\%$.   This led to a systematic uncertainty of $^{+0.30}_{-0.38}$ of the Standard Model cross section.  Four other systematic uncertainties were considered:  uncertainty in the density distribution of the Earth ($\pm 0.01$) \cite{accuracy},   variations in atmospheric pressure at the neutrino production sites ($^{+0.00}_{-0.04}$) \cite{Sandydissertation}, uncertainties in the prompt and astrophysical neutrino spectral indices ($\pm 0.10$), and uncertainties in the angular acceptance of the IceCube DOMs ($^{+0.04}_{-0.00}$).  These systematic errors are then added in quadrature to the systematic uncertainties from the fit, 
giving a total systematic uncertainty of $^{+0.39}_{-0.43}$ times the prediction of the Standard Model.  
 
\begin{table*}[t]
\begin{tabular}{llcrr}
Result& Baseline/units   & Nuisance Parameter & Nuisance Parameter \\
        &            &  Input $\&$ uncertainty $\sigma$       & Fit result  \\
         \hline
$\Phi_{\rm Conv.}\times\sigma$ & Ref.  \cite{Honda:2006qj} $\times  R$ ($R=\sigma_{\rm meas.}/\sigma_{\rm SM}$) &  $1.0 \pm 0.25$      & $0.92 \pm 0.03$ \\
 $\Phi_{\rm Conv.}$  spectral index    &   Ref. \cite{Honda:2006qj} with knee   &   $0.00 \pm 0.05$          &  $+ 0.007\pm 0.001$ \\
 K/$\pi$ ratio				     &  Ref. \cite{Honda:2006qj} baseline               & $1.0\pm 0.1$           & $1.05\pm0.09$ \\
 $\nu/\overline\nu$ ratio                 &    Ref. \cite{Honda:2006qj} baseline               & $1.0\pm 0.1$       & $1.01 \pm 0.005$   \\
$ \Phi_{\rm prompt}\times\sigma$   & Ref.  \cite{Enberg:2008jm}$\times R$           & $0.0_{-0.0}^{+1.0}$ & $0.5^{+0.40}_{-0.34}$ \\
 $\Phi_{\rm astro.}\times\sigma$ &     Ref. \cite{Aartsen:2015knd}$\times R$         & $2.23\pm 0.4$  & $2.62_{-0.07}^{+0.05}$ \\
 Astrophysical index ($\gamma$)    &                                                                         & $2.50\pm 0.09$   &  $2.42\pm 0.02$ \\
 DOM Efficiency		               & IceCube Baseline                                                  & $1.0\pm 0.1$     & $1.05\pm 0.01$ \\
\hline
 \end{tabular}
 \caption{Fit parameters with their baseline or units (second column), along with the prior assumption (initial value) and uncertainty input to the fit (third column) and the values returned by the fit (last column).   The neutrino fluxes are for $\nu_\mu$ and $\overline\nu_\mu$ only. For the astrophysical component, the baseline flux is $\Phi_{\rm astro.}\times(E_\nu/100\ {\rm TeV})^\gamma\ 10^{-18}$
  s$^{-1}$ cm$^{-2}$ sr$^{-1}$.  The three flux terms are multiplied by $R$ to remove the obvious correlation that the number of observed events increases linearly with the cross section, even in the absence of absorption.}
 \end{table*}

The neutrino energy range in which this analysis is sensitive was found by repeating the fit procedure with the absorption probability set to zero for neutrino energies below a certain threshold.  As the threshold was gradually increased, the data and simulation diverged, and the fit quality degraded.    The threshold that corresponded to a likelihood increase of 1.0 $\sigma$ ($-2\Delta$\ LLH = 1, where $LLH$ is the natural log of the likelihood) was the minimum energy for which this analysis was sensitive.    We repeated the process by turning off neutrino absorption above a gradually decreasing high-energy threshold to find the upper end of the energy range.    This procedure gives an energy range of 6.3 TeV to 980 TeV.  This wide range reflects the combination of a neutrino flux that decreases rapidly with energy (partially compensated by an increasing cross section and detection probability) with the relatively rapid increase in absorption with increasing energy. 

 Figure \ref{fig:summary} compares this measurement with previous measurements of neutrino cross sections made at accelerator facilities.  This is the first cross section measurement at multi-TeV energies, where the effects of the finite $W^\pm$ and $Z^0$ masses slow the growth of the cross section with increasing energy.  We measure the cross section to be $1.30^{+0.21}_{-0.19}$ (stat.) $^{+0.39}_{-0.43}$ (syst.)  times the prediction of the Standard Model for charged and neutral current interactions, in the energy range from 6.3 TeV to 980 TeV.     We do not see a dramatic increase in cross section, as is expected in models of beyond-Standard-Model physics, such as those involving extra dimensions \cite{AlvarezMuniz:2002ga} or leptoquarks \cite{Romero:2009vu}.
 
Future measurements with IceCube or larger detectors like IceCube-Gen2 \cite{Aartsen:2014njl} or Phase 2.0 of KM3NeT \cite{Adrian-Martinez:2016fdl} should be able to extend this measurement to higher energies and study the energy-dependence of the cross section.  Future experiments that detect the radio emission from neutrino showers over volumes exceeding 100 km$^3$ \cite{Barwick:2014pca,Allison:2015eky} should observe the interactions of `GZK' neutrinos and extend the cross section measurements up to $10^{19}$~eV \cite{Klein:2013xoa}. These experiments should have extended sensitivity to beyond-Standard-Model processes.  

We acknowledge the support from the following agencies:
U.S. National Science Foundation-Office of Polar Programs,
U.S. National Science Foundation-Physics Division,
University of Wisconsin Alumni Research Foundation,
the Grid Laboratory Of Wisconsin (GLOW) grid infrastructure at the University of Wisconsin - Madison, the Open Science Grid (OSG) grid infrastructure;
U.S. Department of Energy, and National Energy Research Scientific Computing Center,
the Louisiana Optical Network Initiative (LONI) grid computing resources;
Natural Sciences and Engineering Research Council of Canada,
WestGrid and Compute/Calcul Canada;
Swedish Research Council,
Swedish Polar Research Secretariat,
Swedish National Infrastructure for Computing (SNIC),
and Knut and Alice Wallenberg Foundation, Sweden;
German Ministry for Education and Research (BMBF),
Deutsche Forschungsgemeinschaft (DFG),
Helmholtz Alliance for Astroparticle Physics (HAP),
Initiative and Networking Fund of the Helmholtz Association,
Germany;
Fund for Scientific Research (FNRS-FWO),
FWO Odysseus programme,
Flanders Institute to encourage scientific and technological research in industry (IWT),
Belgian Federal Science Policy Office (Belspo);
Marsden Fund, New Zealand;
Australian Research Council;
Japan Society for Promotion of Science (JSPS);
the Swiss National Science Foundation (SNSF), Switzerland;
National Research Foundation of Korea (NRF);
Villum Fonden, Danish National Research Foundation (DNRF), Denmark and the
United States Air Force Academy.

\end{document}